\documentclass[manuscript=fullresearchpaper,american,latin1]{beilstein}

\usepackage{xspace}

\begin{document}

\title{Nonlinear thermoelectric effects in high-field superconductor-ferromagnet tunnel junctions.}
\author{S. Kolenda}
\affiliation{Karlsruher Institut f\"ur Technologie (KIT), Institut f\"ur Nanotechnologie, P.O. Box 3640, D-72021 Karlsruhe, Germany}
\author{P. Machon}
\affiliation{Department of Physics, University of Konstanz, D-78457 Konstanz, Germany}
\author*[1]{D. Beckmann}{detlef.beckmann@kit.edu}
\author*[2]{W. Belzig}{wolfgang.belzig@uni-konstanz.de}
\maketitle

\begin{abstract}
\background Thermoelectric effects result from the coupling of charge and heat transport, and can be used for thermometry, cooling and harvesting of thermal energy. The microscopic origin of thermoelectric effects is a broken electron-hole symmetry, which is usually quite small in metal structures, and vanishes at low temperatures.
\results We report on a combined experimental and theoretical investigation of thermoelectric effects in superconductor/ferromagnet hybrid structures. We investigate the depencence of thermoelectric currents on the thermal excitation, as well as on the presence of a dc bias voltage across the junction.
\conclusion Large thermoelectric effects are observed in superconductor/ferromagnet and superconductor/normal-metal hybrid structures. The spin-independent signals observed under finite voltage bias are shown to be reciprocal to the physics of superconductor/normal-metal microrefrigerators.  The spin-dependent thermoelectric signals in the linear regime are due to the coupling of spin and heat transport, and can be used to design more efficient refrigerators.
\end{abstract}


\section{Introduction}

Electrons in classical superconductors are bound in spin-singlet Cooper pairs, whereas ferromagnetic materials prefer parallel spin alignment. In nanoscale hybrid structures made of superconductors and ferromagnets, the competition of these antagonistic spin orders can be exploited to produce 
superconducting spintronics functionality \cite{eschrig2011,linder2015,eschrig2015}. Several promising spintronic effects have been theoretically predicted and subsequently experimentally observed. Examples are the odd-frequency triplet supercurrent \cite{bergeret:2005,khaire:2010,robinson:2010} and fully spin-polarized quasiparticle currents \cite{huertas:2002,huebler:2012,quay2013}. Superconductor/normal-metal hybrid structures can also be used for local electron thermometry and microrefrigeration \cite{giazotto2006,muhonen2012}. Recently, large spin-dependent thermoelectric effects were predicted \cite{machon2013,machon2014,ozaeta2014,kalenkov2014,kalenkov2015} and experimentally observed \cite{kolenda2016} in superconductor/ferromagnet hybrid structures. These thermoelectric effects are linked to a coupling of spin and heat current, a phenomenon which has recently given rise to the field of spin caloritronics \cite{bauer2012}. 

Most previous works have concentrated on the regime of linear response of the electric and thermal currents to the difference in electric potential or temperature. In that case the linear response coefficients -- electrical and thermal conductance, Seebeck and Peltier coefficients -- are related by the famous Onsager symmetry relations. In particular these relate the thermoelectric responses. In terms of practical usage the linear response coefficients are limited to devices with vanishing performance, due to the assumption of linearization in the thermodynamic forces. E.g. the maximal possible Carnot efficiency $\eta_C=|\delta T|/T$ for a given temperature difference $\delta T$ at base temperature $T$ is by definition is much smaller than 1. Hence, a useful thermodynamic machine need to be run at finite power output, in which the linearization might not work anymore.

In this paper, we extend our previous theoretical \cite{machon2013,machon2014} and experimental \cite{kolenda2016} work in a combined experimental and theoretical study of nonlinear thermoelectric effects in superconductor/ferromagnet hybrid structures, and elucidate the relation of thermoelectric currents to superconducting microrefrigerators by generalizing Onsager relations. 


\section{Experiment \& Results}

\begin{figure}
\caption{(a) False-color scanning electron miscroscopy image of one of our samples, together with the measurement scheme. The samples consist of a six-probe tunnel junction between a superconducting aluminum (Al) and a ferromagnetic (Fe) wire, with an overlaid copper (Cu) wire providing additional measurement leads. (b) Scheme of the generation of the linear thermoelectric effect in a FIS junction. (c) Scheme of the generation of the nonlinear thermoelectric effect in a NIS (or FIS) junction.}
\label{fig:sample}
\includegraphics[width=12.3cm,keepaspectratio]{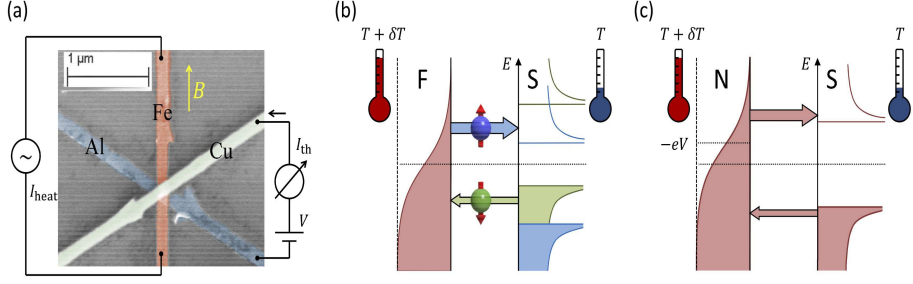}
\end{figure}

Our samples were fabricated by e-beam lithography and shadow evaporation. The central part is a tunnel junction between ferromagnetic iron and superconducting aluminum, with a thin aluminum oxide layer as tunnel barrier. An additional copper wire is overlaid to provide additional measurement leads, forming a six-probe junction. Fig.~\ref{fig:sample}(a) shows a false-color scanning electron microscopy image of one of our samples, together with the measurement scheme. The wire widths are around 200~nm, and the film thicknesses are $t_\mathrm{Al} \approx 20~\mathrm{nm}$, $t_\mathrm{Fe} \approx 15 -20~\mathrm{nm}$ and $t_\mathrm{Cu} \approx 50~\mathrm{nm}$ for the aluminum, iron and copper wires, respectively. Throughout this paper, we will use F, S, I and N to denote ferromagnetic, superconducting, insulating and normal-metal parts of the structures, {\em e.g.}, FIS for a ferromagnet-insulator-superconductor junction.

Transport measurements were carried out in a dilution refrigerator at temperatures down to $50~\mathrm{mK}$, with an applied in-plane magnetic field $B$ along the iron wire. To create a temperature difference $\delta T$ across the junction, we pass a heater current $I_\mathrm{heat}$ along the ferromagnetic wire. The local temperature of the ferromagnet at the junction can be described by \cite{giazotto2006}
\begin{equation}
 T_\mathrm{F}=\sqrt{T^2+\frac{I_\mathrm{heat}^2R_\mathrm{heat}^2}{4L_0}},
   \label{eqn_heatmodel}
\end{equation}
where $T$ is the electronic base temperature without heating, $R_\mathrm{heat}$ is the resistance of the ferromagnetic wire, and $L_0=\pi^2k_\mathrm{B}^2/3e^2$ is the Lorenz number. We calibrate the dependence of $T_\mathrm{F}$ on $I_\mathrm{heat}$ by measuring the differential conductance of the junction while applying a dc heater current. The actual temperature difference $\delta T$ is usually slightly smaller than $\delta T_\mathrm{F}=T_\mathrm{F}-T$ obtained from the calibration measurements due to indirect heating of the superconductor. We typically find $\delta T\approx 0.8\delta T_\mathrm{F}$. Details of the temperature calibration can be found in \cite{kolenda2016}.  

The charge current $I^c$ through a tunnel junction in the presence of a voltage $V$ and a temperature difference $\delta T$ across the junction can be conveniently described in the linear regime by
\begin{equation}
 I^c = gV+\eta \frac{\delta T}{T}
   \label{eqn_Igeneral}
\end{equation}
where $g$ is the conductance, $T$ is the average temperature, and $\eta$ describes the thermoelectric current. $\eta$ is related to the Seebeck coefficient $S=-V/\delta T$ measured in an open circuit by $\eta=SgT$. The physics of the thermoelectric current generation in a high-field FIS junction at zero voltage bias is shown schematically in Fig.~\ref{fig:sample}(b). The Zeeman splitting of the quasiparticle states in the superconductor leads to a spin-dependent density of states (left). Heating of the ferromagnet leads to a flow of spin-up electrons at positive energy from occupied states in the ferromagnet into the superconductor, and a flow of spin-down electrons out of the superconductor into unoccupied states in the ferromagnet at negative energies (relative to the chemical potential of the superconductor). For finite spin polarization $P$ of the junction conductance, the two currents are unequal, and therefore a net charge current flows across the junction, accompanied by both spin and heat currents. Due to the energy dependence of the density of states in the superconductor, the thermoelectric current is a nonlinear function of the thermal excitation $\delta T$.

In the nonlinear regime, the thermoelectric coefficient can be generalized to 
\begin{equation}
 \eta(V,\delta T) = T\left.\frac{\partial I^c}{\partial \delta  T}\right|_{\delta T,V}.
   \label{eqn_Inonlinear}
\end{equation}
At finite voltage bias $V$, schematically depicted in Fig.~\ref{fig:sample}(c), the current through a NIS or FIS junction always depends on temperature, as the forward and backwards currents are always unequal. In this case the generalized nonlinear coefficient $\eta$ is nothing but the temperature dependence of the regular tunnel current.

In our previous work \cite{kolenda2016}, we focussed on the measurement of $\eta$ for a fixed thermal excitation $\delta T$ at $V=0$. Here, we elucidate the nonlinear regime with data for different $\delta T$ and finite voltage bias $V$. To measure the thermoelectric current through the junction, we apply a low-frequency ac heater current. Since the heating power is proportional to $I^2$, this generates a thermal excitation on the second harmonic of the excitation frequency. We monitor the second harmonic of the current $I_\mathrm{th}$ through the junction, which is proportional to the nonlinear coefficient $\eta$ given by Eq.~(\ref{eqn_Inonlinear}). We show data from three samples, two with ferromagnetic junctions (FIS1 and FIS2), and a reference sample where the iron wire is replaced by copper to form a nonmagnetic junction (NIS). Details of the sample parameters and characterization can be found in \cite{kolenda2016}.

\begin{figure}
\caption{(a) Thermoelectric current $I_\mathrm{th}$ as a function of thermal excitation amplitude $\delta T$ for different magnetic fields $B$ (sample FIS1). (b) thermoelectric transport coefficient $\eta$ normalized to $G_\mathrm{T} \Delta_0/e$ corresponding to the data in panel (a).}
\label{fig:nl_amplitude}
\includegraphics[width=8.2cm,keepaspectratio]{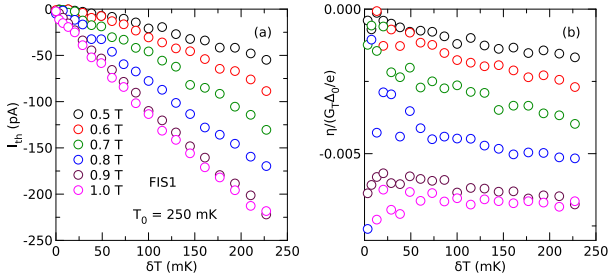}
\end{figure}

In Fig.~\ref{fig:nl_amplitude}(a), we show the thermoelectric current $I_\mathrm{th}$ as a function of thermal excitation $\delta T$ for different magnetic fields $B$ at a base temperature $T_0=250~\mathrm{mK}$ measured in sample FIS1. 
The maximum current is observed around $B=1~\mathrm{T}$. At this field, the spectral gap of the superconductor vanishes (see Figure~2(a) of \cite{kolenda2016}), and the thermoelectric current is a nearly linear function of the excitation. At smaller fields, the superconductor has an energy gap, and as a consequency the thermoelectric current is smaller, and has a nonlinear dependence on the excitation. In Fig.~\ref{fig:nl_amplitude}(b), we show the corresponding thermoelectric coefficient $\eta=T I_\mathrm{th}/\delta T$, normalized to $G_\mathrm{T} \Delta_0/e$, where $G_\mathrm{T}=275~\mathrm{\mu S}$ is the normal-state junction conductance, and $\Delta_0=208~\mathrm{\mu eV}$ is the pair potential of the superconductor at $T=0$ and $B=0$. $\eta$ is nearly constant at high fields, and has a weak dependence on the excitation $\delta T$ at smaller fields.

\begin{figure}
\caption{Thermoelectric transport coefficient $\eta$ normalized to $G_\mathrm{T} \Delta_0/e$ as a function of bias voltage $V$ for different applied magnetic field $B$. (a) data for a ferromagnetic junction (sample FIS2). (b) data for a nonmagnetic junction (sample NIS).}
\label{fig:nl_bias}
\includegraphics[width=8.2cm,keepaspectratio]{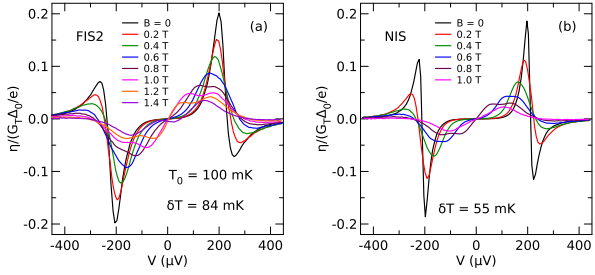}
\end{figure}

In Fig.~\ref{fig:nl_bias}, we compare the nonlinear thermoelectric coefficient $\eta$ for two samples, one with a ferromagnetic junction (a), and one with a normal-metal junction (b). $\eta$ is plotted as a function of voltage bias $V$ for fixed thermal excitation $\delta T$ at different magnetic fields. While the nonmagnetic sample does not show a linear thermoelectric effect (see also Fig.~4(c) of \cite{kolenda2016}), both samples show a large nonlinear effect, even at zero applied field. Note that the overall signal scale is about two orders of magnitude larger than in Fig.~\ref{fig:nl_amplitude}. The linear thermoelectric effect at $V=0$, which appears exclusively in the FIS sample, is hardly visible on this scale due to the small spin polarization $P=0.08$ of our samples. The nonlinear coefficient is an odd function of bias, and has a sign reversal for bias voltages close to the energy gap of the superconductor.

\section{Theory}

In the linear response regime the Seebeck and the Peltier coefficients are related by the Onsager reciprocity relation. Hence a measurement of one determines the other. This is not the case in the nonlinear regime anymore. In the following we derive a generalization of the Onsager relation in the nonlinear regime 
to evaluate the performance of mesoscopic cooling devices. Obviously this cannot be as general as the Onsager reciprocity, but relies on a conrete model of elastic transport. In the end it  will be useful to evaluate the practically important heat current from the measure thermally induced charge current.

We consider a metal coupled to a superconductor by a tunnel contact. The metal can be a normal metal or a ferromagnet. In that context the superconductor is kept at zero chemical potential. We can in general express the charge and heat current as
\begin{align}
	I^c(V,\delta T) & =  \int \frac{dE}{e} G(E) \left( f_{T+\delta T}(E-eV)-f_T(E) \right)\\
	I^Q(V,\delta T) & = I^E(V,\delta T)-VI^c(V,\delta T) \\
		& = \int dE \frac{G(E)}{e^2} (E-eV) \left( f_{T+\delta T}(E-eV)-f_T(E) \right)\label{eqn:IQ}
\end{align}
Here $G(E)$ is the spectral conductance and $f_T(E)=(\exp(E/k_BT)+1)^{-1}$ is the Fermi function at energy $E$. Note that we assume the spectral conductance to be independent of temperature and bias voltage. This is in general not always fulfilled, since e.g. the superconducting gap $\Delta$ depends on temperature. However this becomes mainly relevant close to $T_c$ and we will in the following neglect the temperature dependence. The following derivation will be based on the identity
\[
	\left.\frac{\partial}{\partial \delta T}f_{T+\delta T}(E-eV)\right|_{\delta T=0}
	=\frac{E-eV}{4k_BT^2}\frac{1}{\cosh^2\frac{E-eV}{2k_BT}}
	=\frac{E-eV}{T}\frac{\partial}{\partial eV}f_T(E-eV)
\]
valid for arbitrary bias voltage. Hence we can write
\begin{align}
	\frac{\partial I^Q(V,0)}{\partial V} & = \frac{\partial I^E(V,0)}{\partial V}-V\frac{\partial I^c(V,0)}{\partial V}-I^c(V,0) \\
		& = \int \frac{dE}{e^2} G(E) (E-eV) \frac{\partial}{\partial V}f_T(E-eV) -I^c(V,0)\\
		& = \int  \frac{dE}{e}G(E) T\left.\frac{\partial}{\partial \delta T}f_{T+\delta T}(E-eV)\right|_{\delta T=0} -I^c(V,0)\\
		& = T\left.\frac{\partial}{\partial \delta T}I^c(V,\delta T)\right|_{\delta T=0}-I^c(V,0),
\end{align}
and finally
\begin{equation}
  \frac{\partial I^Q(V,0)}{\partial V} = \left.\eta(V,\delta T)\right|_{\delta T=0}-I^c(V,0).\label{eqn:analysis}
\end{equation}

\begin{figure}
\caption{(a) Normalized cooling power $I^Qe^2/G_\mathrm{T}\Delta_0^2$ as a function of normalized bias voltage $eV/\Delta_0$ for different magnetic fields $B$. (b) Predicted cooling power for the same device assuming $P=0$ (NIS cooler) and $P=1$ (ideal FIS Peltier cooler) as a function of normalized bias voltage. (c) Predicted coefficient of performance as a function of normalized cooling power for the same parameters as panel (b) and $V<0$.}
\label{fig:pow}
\includegraphics[width=12.3cm,keepaspectratio]{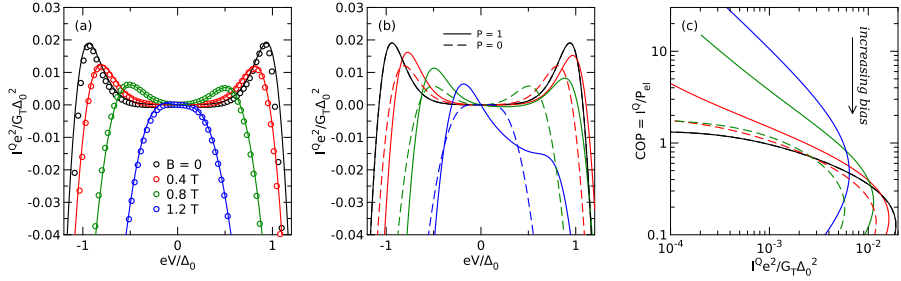}
\end{figure}

This is the main result and can directly be applied to the experimental data. In Fig.~\ref{fig:pow}(a), we show the cooling power $I^Q$ predicted from the measured thermoelectric coefficient $\eta$ and dc current $I^c$ of sample FIS2, using Eq.~(\ref{eqn:analysis}) and integrating over $V$. Symbols are experimental data, while lines are fits using Eq.~(\ref{eqn:IQ}) directly. The data and fits are in good agreement, showing that the cooling power can be reliably predicted from the measured thermoelectric coefficient in the nonlinear regime. At $B=0$, without spin splitting and consequently without linear thermoelectric effect, the predicted cooling power has the typical bias dependence of NIS microrefrigerators \cite{giazotto2006}, with maximum cooling power for $eV\approx \Delta$. Upon increasing the field, the maximum of the cooling power shifts to smaller bias and decreases. Note that the Peltier cooling at zero bias due to the linear thermoelectric effect is too small to be resolved in this plot due to the low spin polarization $P=0.08$ of our junction. Using the sample parameters of the fits shown in Fig.~\ref{fig:pow}(a), we can now compare the predicted cooling power of a NIS cooler and an idealized FIS cooler with $P=1$ in Fig.~\ref{fig:pow}(b). As can be seen, there is no difference between NIS and FIS at $B=0$. At finite field, the FIS cooler exhibits a linear Peltier contribution to the cooling power, which is largest at $B=1.2~\mathrm{T}$, roughly where the gap in the excitation spectrum of the superconductor vanishes. Under these conditions, the FIS Peltier cooler outperforms the NIS cooler at small bias. It is convenient to define the coefficient of performance $COP$ for a cooler as the ratio $COP=I^Q/P_\mathrm{el}=I^Q/I^cV$ of the cooling power and the electric input power of the device \cite{whitney2014}. To make the improved performance of the FIS cooler more clear, we also plot the coefficient of performance as a function of cooling power in Fig.~\ref{fig:pow}(c). The FIS cooler has superior efficiency over a wide range of cooling powers.

\section{Discussion}

The thermoelectric current is largest and has a linear dependence on excitation at the magnetic field where the spectral gap of the superconductor vanishes. These conditions are therefore potentially useful for applications in thermometry or cooling. One possible way to improve performance is therefore to increase the spin splitting of the density of states by spin-active scattering with a ferromagnetic insulator \cite{moodera1988,hao1991}, which is known to enhance nonequilibrium spin transport in nanoscale superconductors \cite{wolf2014c}. Also, performance can be improved by using ferromagnetic insulators as spin-filter tunnel junctions, with a degree of spin polarization $P\approx 100\%$ \cite{hao1990,miao2015}.

At finite voltage bias, we find large thermoelectric signals for both FIS and NIS structures. Our analysis based on a generalized reciprocity relation shows that the generation of the thermoelectric signal is directly related to the cooling power of NIS microrefrigerators \cite{giazotto2006,muhonen2012}. Further theoretical modeling shows that for an idealized FIS cooler with $P=100\%$, the thermodynamic efficiency can be greatly improved over NIS coolers. Future devices may include  local control of the spin-splitting using the proximity effect with ferromagnetic insulators \cite{wolf2014b,wolf2014c}, or new thermoelectric multi-terminal devices \cite{machon2013,machon2014}.


\begin{acknowledgements}
We acknowledge financial support by the competence network ``Functional Nanostructures'' of the Baden-W\"urttemberg-Stiftung and the DFG under grant No. BE-4422/2-1 and BE-3803/3-1.
\end{acknowledgements}

\bibliography{KFN_C2}

\end{document}